# Prediction of motion-induced magnetic fields for human brain MRI at 3T

Jiazheng Zhou[1,2], Gisela E Hagberg[1,3], Ali Aghaeifar[1,3], Jonas Bause[1], Maxim Zaitsev[4], Klaus Scheffler[1,3].

1. High-Field Magnetic Resonance Center, Max Planck Institute for Biological Cybernetics, Tübingen, Germany
2. IMPRS for Cognitive and Systems Neuroscience, University of Tübingen, Tübingen, Germany
3. Department of Biomedical Magnetic Resonance, University Hospital Tübingen, Tübingen, Germany
4. Department of Radiology, Medical Physics, Medical Center University of Freiburg, Faculty of Medicine, University of Freiburg, Freiburg, Germany

Corresponding author: Jiazheng Zhou
Address: High-Field Magnetic Resonance Center, Max Planck Institute for Biological Cybernetics, Max-Planck-Ring 11, 72076 Tübingen, Germany.
Email: jiazheng.zhou@tuebingen.mpg.de

**Abstract**

Objective

Maps of $B_0$ field inhomogeneities are often used to improve MRI image quality, even in a retrospective fashion. These field inhomogeneities depend on the exact head position within the static field but acquiring field maps (FM) at every position is time consuming. Here we explore different ways to obtain $B_0$ predictions at different head-positions.

Methods

FM were predicted from iterative simulations with four field factors: 1) sample induced $B_0$ field, 2) system's spherical harmonic shim field, 3) perturbing field originating outside the field-of-view, 4) sequence phase errors. The simulation was improved by including local susceptibility sources estimated from UTE scans and position-specific masks. The estimation performance of the simulated FMs and a transformed FM, obtained from the measured reference FM, were compared with the actual FM at different head positions.

Results

The transformed FM provided inconsistent results for large head movements (>5° rotation), while the simulation strategy had a superior prediction accuracy for all positions. The simulated FM was used to optimize $B_0$ shims with up to 22.2% improvement with respect to the transformed FM approach.

Conclusion

The proposed simulation strategy is able to predict movement-induced $B_0$ field inhomogeneities yielding more precise estimates of the ground truth field homogeneity than the transformed FM.

# 1. Introduction:

Spatial encoding and signal acquisition in Magnetic resonance imaging (MRI) rely on a homogenous static magnetic field ($B_0$). Spatially varying magnetic susceptibilities within samples can perturb the static magnetic field. These perturbations are prominent at the boundaries of tissues with distinct susceptibilities. In particular, significant perturbative effects between anatomic air cavities and biological tissue borders can lead to local frequency deviations and reduce the quality of magnetic resonance measurements. For example, $B_0$ field inhomogeneity causes signal voids and geometric distortion in gradient recalled echo-echo planar images (GRE-EPI)[1]. Signal loss due to intravoxel dephasing can be partly recovered through the use of acquisition-based methods, such as Z shimming[2] or hybrid RF pulse design[3]. The geometric distortion artifact could be corrected or substantially reduced in images through post-processing[4] or $B_0$ shimming[5], provided that knowledge of the field distribution is available.

Functional MRI (fMRI) studies heavily rely on $B_0$ field homogeneity. The experiments can last several minutes, during which the Echo Planar Imaging (EPI) sequence is used to continuously capture the dynamic blood-oxygenation-level-dependent (BOLD) signal. The EPI volumes in presence of $B_0$ inhomogeneity are distorted due to the long acquisition times required to traverse the k-space along the phase-encode direction, but retrospective correction methods based on a field-map measurement can be used to improve image quality[6]. $B_0$ shim coils are devices that improve $B_0$ homogeneity by providing a time invariant and spatially smooth magnetic field that counter-acts $B_0$ inhomogeneity caused by the measured object. In practice, studies using whole-body spherical harmonic (SH) shimming[7] and localized multi-coil shimming[8–10] have demonstrated the impact of improving $B_0$ field homogeneity on the quality of EPI images. Both retrospective and prospective $B_0$ shimming methods require knowledge of $B_0$ the field distribution, whereas the field map is typically obtained once prior to (or after) the experiments. Long-lasting experiments (e.g., fMRI) are known to be prone to subject head movements. The head motion alters the position and orientation of the susceptibility interfaces in relation to the static $B_0$ field and accordingly alters the inhomogeneity distribution. Therefore, the measured field map acquired only once may not be valid for the correction of geometric distortions for the entire fMRI session. A typical whole-brain field map acquisition with dual-echo GRE takes 1-3 minutes depending on the field-of-view (FOV), resolution, and acceleration provided by parallel imaging. Due to time constraints, it may be impractical to repeat it several times for each and every head position. As an alternative, field maps can be calculated from phase offsets of two EPI measurements[11, 12]. However, this requires a sequence modification to implement jittered echo time[13]. Additionally, sequence can be modified to implement methods like 3D EPI navigator[14] and FID navigator[15] for rapid estimation of the field map.

In principle, if all susceptibility sources, including their shape, and orientation with respect to the external field were known, the magnetic field map could be obtained from analytic magnetostatic equations[16]. However, analytical solutions are only practical for simplified geometries (like spheres or cylinders), making this approach very difficult to be used with complex structures. Alternatively, rapid macroscopic dipole approximations methods[17, 18] give a numerically approximated solution of susceptibility induced $B_0$ field inhomogeneities. This allows the estimation of $B_0$ field inhomogeneities over arbitrary sample structures and significantly reduces computational time. The mentioned computational approach requires an accurate construction of susceptibility models specific for each sample. When imaging a phantom, such models can be constructed based on the design data and material properties[19]. In the human head, the primary susceptibility components are the brain, bone, and air. For accurate modeling of field inhomogeneities, it is most critical to differentiate 1) air/bone air/tissue cavity boundaries and 2) their susceptibility values. However, none of these cavity boundaries yield detectable signals with conventional MRI sequences. Therefore, initial studies were based on the fusion of computed tomography (CT) and MRI images to make such a head model[20, 21]. Although the susceptibility model is helpful, it is not subject-specific and depends on the co-registration process to accurately localize the sinuses in MR images. One possibility to observe and separate air/bone boundaries using MR images only, is by utilizing an ultra-short echo time (UTE) sequence which is suitable for the detection of signals from tissues with very short T2 components (e.g., bone) with nominal TEs (<< 1ms). A similar approach using a dual-echo short TE 3D GRE sequence for air-tissue boundary segments has provided promising results[22] for background field removal applications such as susceptibility-weighted imaging (SWI)[22] and quantitative susceptibility mapping (QSM)[23]. Such simplified models may not reflect the full complexity of the B0 field generated by the human head.

Here, we propose a field map simulation approach to predict the $B_0$ field at different head positions. It encompasses both global and more local effects by utilizing a susceptibility model built from UTE scans. We systematically identify four field factors: 1) sample induced $B_0$ field, including local susceptibility sources; 2) system's spherical harmonic (SH) shim field, 3) perturbing field originating outside the Field-of-View, and 4) phase errors caused by gradient timing offset while the MR sequence is played out. Comparisons of simulated field maps with ground-truth measured field maps at different head positions were performed. We also compared the estimation performance of the simulated FMs with FMs obtained after spatial transformation of the FM measured at the reference position. Finally, we used the simulated FM to obtain $B_0$ shim currents. We found that the simulated field map can be considered as a potential substitute for a measured field map when this is not available.

**2. Method:**
2.1 Experiment
All measurements were performed on a Siemens Prisma Fit 3 Tesla scanner (Siemens Healthineers, Erlangen, Germany). Scans were performed by using the scanner body coil for RF transmission and the 64-channel head array for signal reception. Four healthy volunteers (3 male and 1 female) were recruited and provided a written informed consent after a full explanation of the protocol. The study was conducted in accordance with the University Hospital of Tuebingen Ethics Review Board.

Experiments were carried out by measuring the volunteer's structural images forming for head model and field maps at four head positions. In the first head position, 2$^{nd}$ order SH shim for the whole head and neck region and frequency adjustment was performed followed by a series of sequences, including: 1) Dual-echo 3D GRE sequence to measure reference B$_0$ maps for the subsequent calculations (TE$_{1/2}$ = 2.68/7.49 ms, TR = 11 ms, FA = 12°, FOV = 256*256*192 mm$^3$, voxel size = 1 mm$^3$ isotropic, GRAPPA factor = 2, monopolar readout gradients and "whisper" gradient mode, which reduces the slew rate, were used to minimizing possible eddy currents); 2) 3D stack-of-spiral UTE sequence for imaging short T2 anatomical components (e.g., bone) (Siemens WIP 992D, TE = 0.05 ms, TR = 8 ms, FA = 1°, FOV = 256*256*192 mm$^3$, voxel size = 1 mm$^3$ isotropic). Later, the volunteers were asked to rotate their head about the Z axis (Position 2), the X axis in both opposite directions (Position 3, Position 4). The dual-echo 3D GRE was repeated to measure reference B$_0$ maps at each head position Position 2, Position 3, and Position 4. System's SH shim parameters were recorded for the simulation. The SH shim remained the same during the entire experiment, in order to imitate the SH shimming conditions where the volunteer moves during a continuous scan.

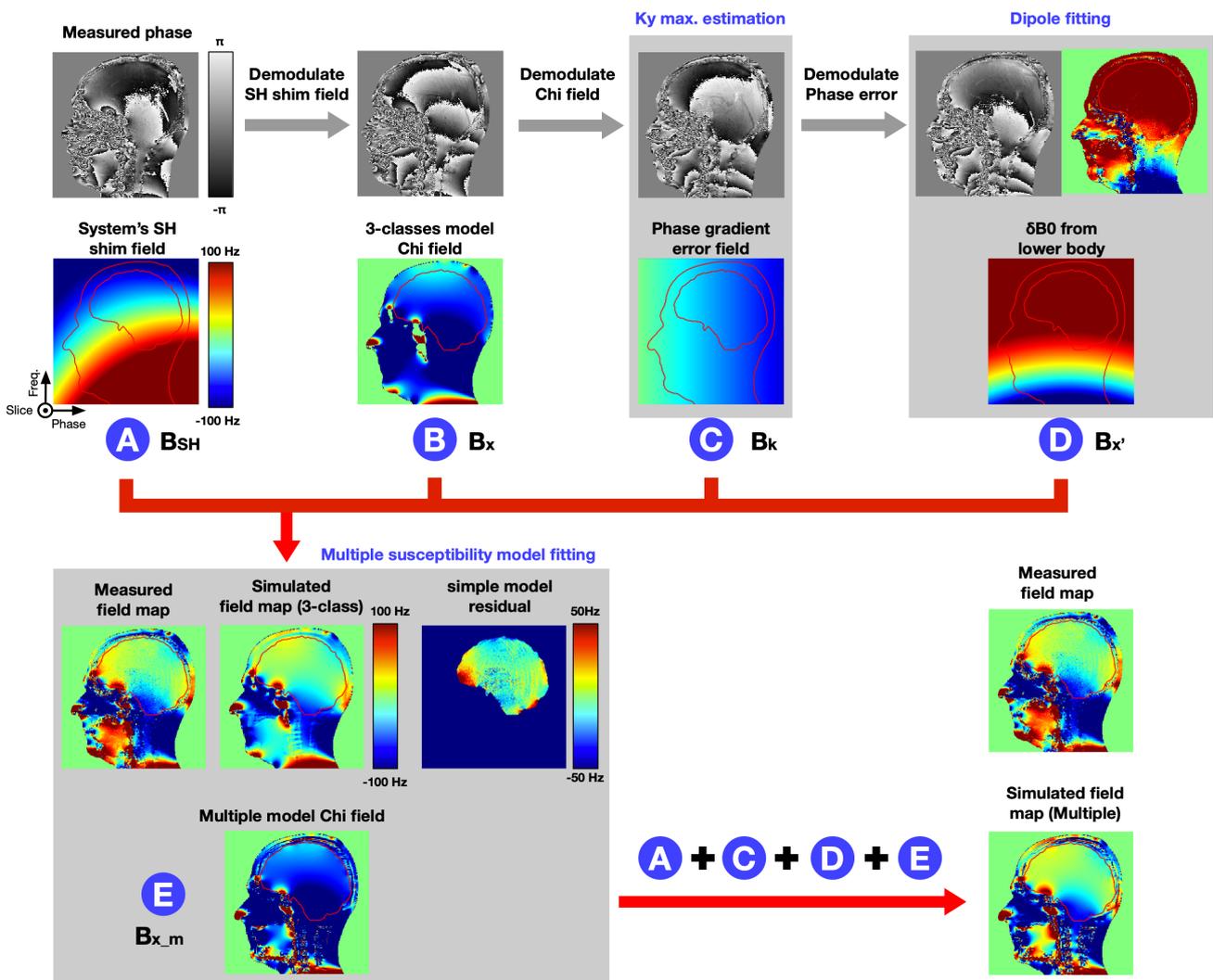

**Fig. 1** *Schematic of simulated field map. The simulation field map is based on a 5 steps magnetic field estimation: (A) the system SH shim field is calculated from the SH shim coefficients; (B) the forward approximated sample induced $B_\chi$ field; (C) Linear phase errors by the imaging gradients; (D) Dipole approximation of the $B_{\chi'}$ field from the lower body part; (E) Susceptibility model fitting to reduce the simulation residual error with the $B_{\chi\_m}$ field.*

2.2 Simulated field map

The deviation ($\Delta B_0$) from a homogeneous magnetic field can be approximated as a superposition of 1) The static magnetic field, $B_0$ (on the order of Tesla). The static magnetic field $B_0$ is commonly treated as a constant; however, in scans with large FOV (e.g., upper limb) or at large off-isocenter locations the static magnetic may not be assumed to be constant across the entire imaging volume. We omitted this factor in later considerations since our focus is on the human brain; 2) An inhomogeneous field $B_\chi$ (on the order of ppm of the static field), induced by the samples (e.g., the human head); 3) The static SH shim field generated by the scanner $B_{SH}$ to counteract the magnetic field inhomogeneities due to the sample-induced field $B_\chi$; 4) The chemical shift effects $B_\sigma$ from lipids and other compounds; 5) The field generated by eddy currents of the gradients system and other potential errors in the phase encoding direction $B_k$, during the 3D GRE field mapping. 6) The perturbating field $B_{\chi'}$ originates from the tissue susceptibility sources (e.g., lungs) located outside the FOV.

Thus, the simulated field map $\Delta B_0$ could be described as Equation 1 (Figure 1):

$$\Delta B_0 = B_\chi + B_{SH} + B_\sigma + B_k + B_{\chi'} \qquad (1)$$

The simulation process as well as the subject susceptibility model were developed in MATLAB (MathWorks Inc., Natick, USA).

2.2.1 Forward calculation of $B_\chi$

To rapidly compute magnetic field inhomogeneities over arbitrary sample geometries, we used the Fourier dipole approximation method that was derived by Marques and Bowtell[18] and Salomir[17] et al. This method only requires Fourier transforms over the input susceptibility distributions (describe in the 2.3 Susceptibility model section), with additional consideration of the Lorentz sphere correction term for microscopic susceptibility correction to Maxwell's equations, as described by Equation 2.

$$B_\chi(\boldsymbol{r}) = FT^{-1}\left\{B_0\left[\frac{1}{3} - \frac{k_z^2}{k_x^2 + k_y^2 + k_z^2}\right] \cdot \tilde{\chi}(\boldsymbol{k})\right\} \qquad (2)$$

Here the $\tilde{\chi}$ indicates the 3-dimensional Fourier transform of the susceptibility distribution and **k** indicates the k-space vector. Susceptibility is weighted by a k-space scaling factor (the terms in brackets), which represents the Lorentz sphere corrected dipole response of the system to an external field. The susceptibility-induced magnetic field in image-space is then given by the inverse Fourier transform of the term in curly brackets.

2.2.2 SH shim field ($B_{SH}$) simulation

The SH shim field from the scanner could be either measured on a spherical phantom or simulated from the vendor-provided SH shim coefficients. In a motion experiment, the subject's movement might cause that measured VOI (e.g., brain) to fall outside of the phantom dimensions. Thus, we opted to simulate the SH shim field (up to 2$^{nd}$ order SH) within the imaging FOV. (Figure 1A)

2.2.3. The chemical shifts effect $B_\sigma$ in the brain

As shown in supplementary figure S1, the chemical shift effect plays a minor role in predicting the field in the brain tissue. Therefore, in our simulation, we did not consider any chemical shift effects inside the brain. However, for other body parts (e.g., liver), both chemical shift and sample susceptibility effects are linearly equally dependent on the main static field strength and need to be considered. Conventional phase contrast cannot distinguish between both sources; therefore, chemical shifts can falsify measurements in MRI when not considered properly.

2.2.4 Estimation of phase error $B_k$

Although the dual-echo 3D GRE sequence protocol was optimized to minimize the possible eddy current effects, we still observed the phase error contributions (Figure 1C).

These errors were modeled by linear fitting of the maximal $k_y$ shift of the complex signal in the k-space center for each echo time ($t_n$) after demodulating the effects of SH shim and subject specific $B_\chi$ field from the measured dual-echo 3D GRE sequence (Equation 3).

$$\Delta \boldsymbol{k}_n = -a t_n \tag{3}$$

where, the $\Delta k_n$ is the k-space shift for echo time $t_n$, and **a** is the slope. We set the linear fit intercept **b** to zero, since $\Delta k_0 = t_0 = 0$.

2.2.5 Magnetic dipole fitting of $B_{\chi'}$

The forward calculation method requires a whole sample susceptibility model (e.g., the entire human body), which is typically not available if the sample susceptibility model is from a scan with limited FOV (e.g., human head or brain). But the $\Delta B_0$ field in the smaller FOV will also be affected by a perturbing field $B_{\chi'}$ from the entire sample. Here we approximate the $B_{\chi'}$ field by a single magnetic dipole field, generated by a source located outside of the region of interest[24].

Equation 4 shows the perturbing field $B_{\chi'}$ at position **r**, modelled an axially oriented magnetic dipole. It is given as a function of four parameters, the dipole strength P and position $\boldsymbol{r}_d = x_d \widehat{\boldsymbol{x}} + y_d \widehat{\boldsymbol{y}} + z_d \widehat{\boldsymbol{z}}$, of the dipole.

$$B_{\chi'}(\boldsymbol{r}) = \frac{P}{|\boldsymbol{r} - \boldsymbol{r}_d|^3} \left( 3 \left( \frac{(\boldsymbol{r} - \boldsymbol{r}_d) \cdot \widehat{\boldsymbol{z}}}{|\boldsymbol{r} - \boldsymbol{r}_d|} \right)^2 - 1 \right) \tag{4}$$

The *P* and $\boldsymbol{r}_d$ were determined by minimizing a cost function (Equation 5), which was defined as the root-mean-square error (RMSE) between the $B_{\chi'}$ and $B_{diff} = B_{measured} - B_{SH} - B_\chi - B_k$, using the "fminsearch" function in MATLAB,

$$cost = \sqrt{\frac{1}{n}\sum_{r=1}^{n}(B_{diff}(r) - B_{\chi'}(r) - B_{0\_shift})^2} \quad (5)$$

This method is valid since the field contributions from the other sources/factors have been mostly removed ($B_{diff}$) or significantly smaller, which leaves only a "linear" like perturbating field in the brain (see Figure 1D). A constant $B_{0\_shift}$ factor (to make sure the simulated FM and measured FM has similar zero offset) was estimated, together with the lower body induced Chi field.

2.3 Susceptibility model

Within the imaging FOV, the field contributions, such as $B_{SH}$, $B_k$, and $B_{\chi'}$ are independent factors of susceptibility sources within the brain. Therefore, we used an iterative nonlinear optimization algorithm to fit a more detailed subject-specific susceptibility model.

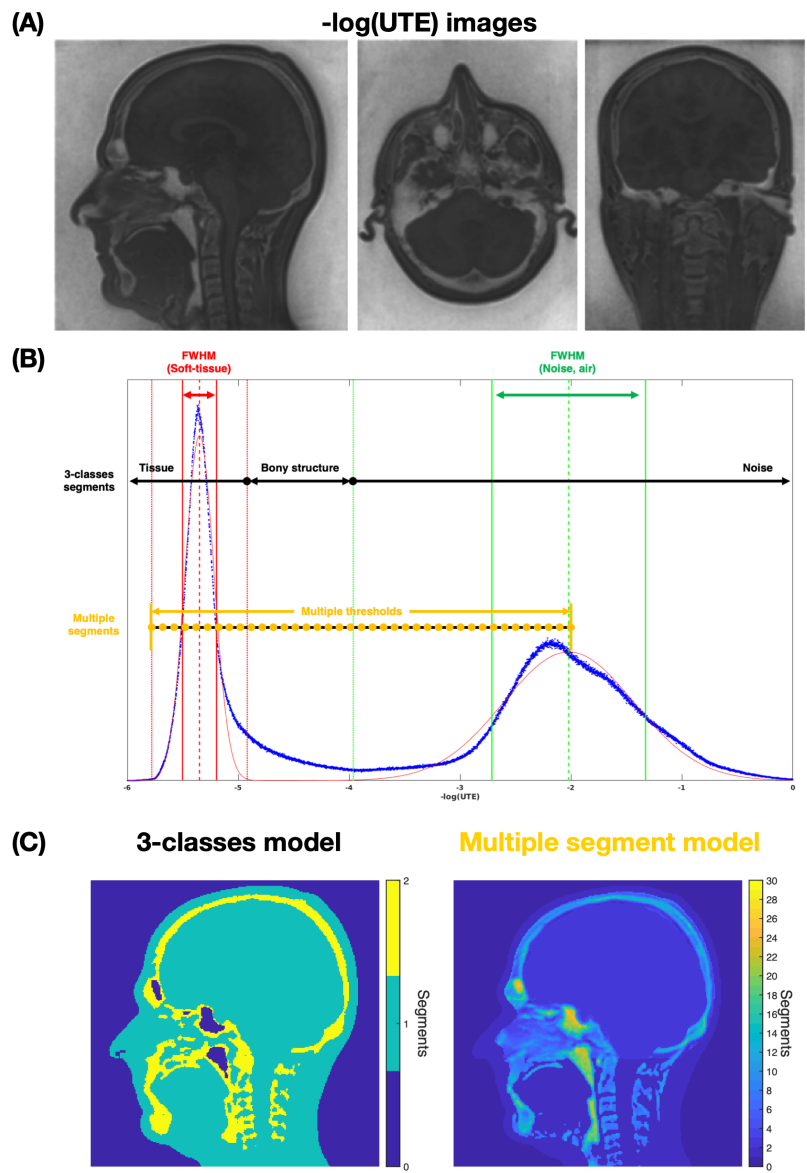

**Fig. 2** *The threshold-based air, bone, and soft tissue segmentation of inverting the logarithmically scaled UTE dataset. (A) Three orthogonal views of the inverted logarithmically scaled UTE image, with full range [-7,0]. (B) The image histogram with two distinctive peaks as soft-tissue (left) and noise (right) signals. Gaussian-fitting results with center peak (dashed line) and full-width-half-maximum (FWHM) information (dashed lines) for soft-tissue (red) and noise (green). The 3-class segment model is based on the bone signal threshold [Center (soft)+1.4\*FWHM (soft), Center (noise)-1.4\*FWHM (noise)], which is positioned in between the two peaks, leading more toward soft tissue. The multiple segment model is defined as a linear interval between [Center (soft)-1.4\*FWHM (soft): 0.1: Center (noise)]. (C) Middle sagittal slices of 3-classes segment model and multiple segment model.*

Tissue compartments, including bone, soft tissue, and cavities were segmented from the proton density weighted UTE image using a threshold-based segmentation method[25]. The UTE image was first bias-corrected for spatial intensity variations using an N4-bias filter[26] (Figure 2A). Then an image histogram was generated after inverting the logarithmically scaled UTE dataset, with soft tissue (left) and air (noise) appearing as two distinct peaks where the peak on the left represents soft tissue (Figure 2B). The bone signals are spread in-between these peaks, trending more toward the soft tissue peak. Two tissue segment models were developed (Figure 2C). A 3-class model where the head was segmented into air, bone, and soft tissue using empirically chosen bone threshold derived from Gaussian fit of the soft-tissue and noise peaks, ranging from [Center (soft)+1.4\*FWHM (soft), Center (noise)-1.4\*FWHM (noise)] (Figure 2B). However, partial volume effects in the nasal area and ear canals often yielded misclassification, since these are formed of complex structures comprising soft tissue, mucus, air, and cartilage, which usually result in high partial volume effects. To better represent these different contributions and correct for that a multiple segments model which was characterized as a linear interval between [Center (soft)-1.4\*FWHM (soft): 0.1: Center (noise)] (Figure 2B) to cover nearly all intermediary tissue classes and cavities. These settings and procedures were reproducible across all participating volunteers

First, a 3-class susceptibility map was determined by assigning typical tissue susceptibility values from the literatures to the three tissue compartments [air = 0 ppm (reference), bone = -11.4 ppm[23], soft tissue = -9.6 ppm[27]]. The 3-class susceptibility map was used to rapidly calculate field contributions of $B_k$, and $B_{\chi'}$ (Figure 1A-D). Compared to the measured field map, the residual field after demodulating field contribution of $B_\chi$, $B_{SH}$, $B_k$, and $B_{\chi'}$ is relatively small (Figure 1E) and local. Thus, we hypothesized this simple model residual field, SMR, is majorly raised from the incorrectly assigned tissue and cavity boundaries, and susceptibility values vary between the individual and the literature when using the 3-class susceptibility map.

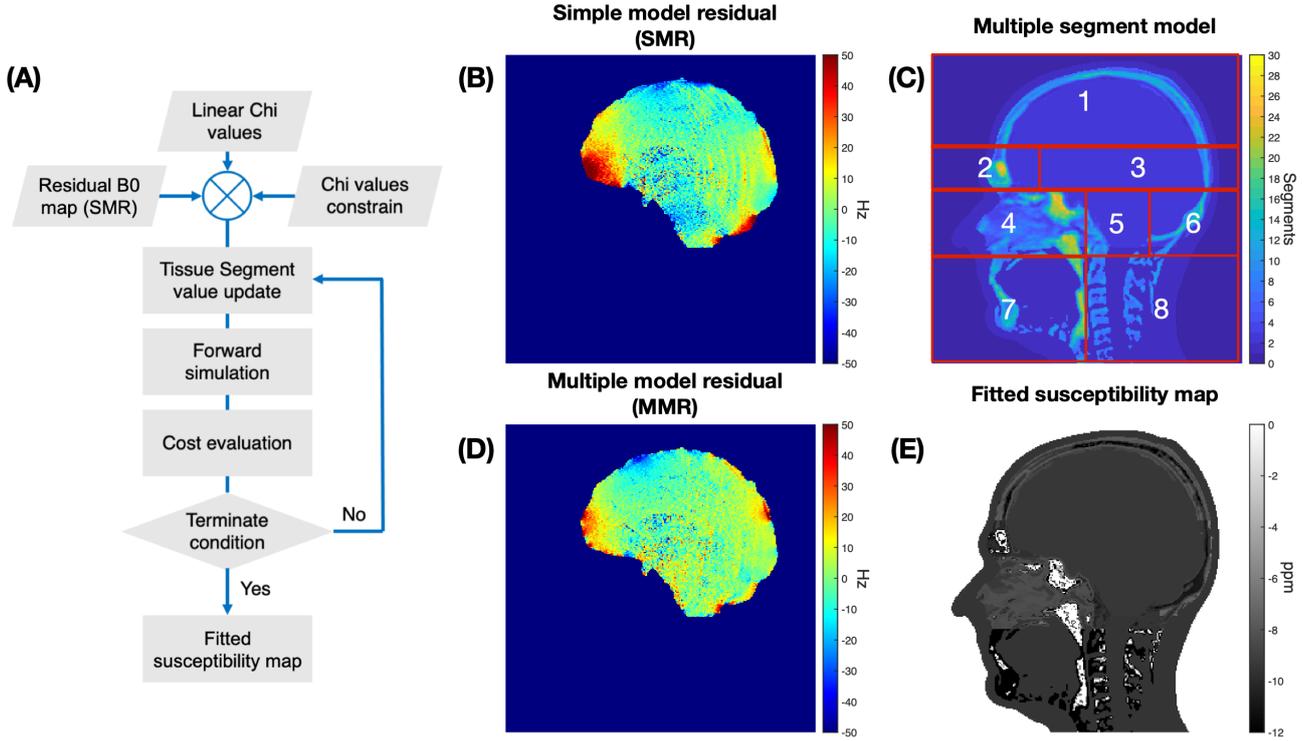

**Fig. 3** *(A) Schematic of the iterative fitting algorithm. (B) The simple model residual field map, SMR, when using a 3-class susceptibility map. (C) The multiple segment model with additional spatial anatomical constrains. (D) The multiple model residual field map, MMR, when using a fitted susceptibility map (E) The fitted susceptibility map for one volunteer.*

Intuitively, by identifying the susceptibility values for each voxel in a tissue segment map could yield a more detailed reasonable $B_{\chi\_m}$ field. We use the "fmincon" function in MATLAB (Figure 1A) to approximate the susceptibility values for multiple tissue compartments model (Figure 3C), through minimizing a cost function (Equation 6) defined as the RMSE between the $B_{\chi\_m}$ field and the $B_{diff\_m} = B_{measured} - B_{SH} - B_{\chi'} - B_k$

$$cost\_m = \sqrt{\frac{1}{n}\sum_{r=1}^{n}(B_{diff\_m}(r) - B_{\chi\_m}(r))^2} \qquad (6)$$

To avoid susceptibility values being constant throughout different anatomical regions (spatially constant), eight volumes of interest (VOIs) were defined to target brain regions with different types of air-tissue interfaces within the multiple segment model (supplementary figure S2): 1) top skull, 2) frontal sinus, 3) middle skull, 4) nasal cavities, 5) ear canals, 6) lower skull, 7) jaw and airway and 8) spine. Between different VOIs, the same tissue compartment should have the same susceptibility values for each region. Additionally, to reduce the number of voxels that required fitting, hence, reducing the computational time, all voxels within the VOIs in the jaw and neck regions that could be assigned to the brain or muscle tissues were assigned the susceptibility values from the literature (Figure 3E). The lower and upper bounds for the susceptibility value were [-14ppm, 1ppm]. For the initial starting point, we used a linear susceptibility interval of [-12ppm, 0ppm], where the incremental step was calculated to match the segment classes.

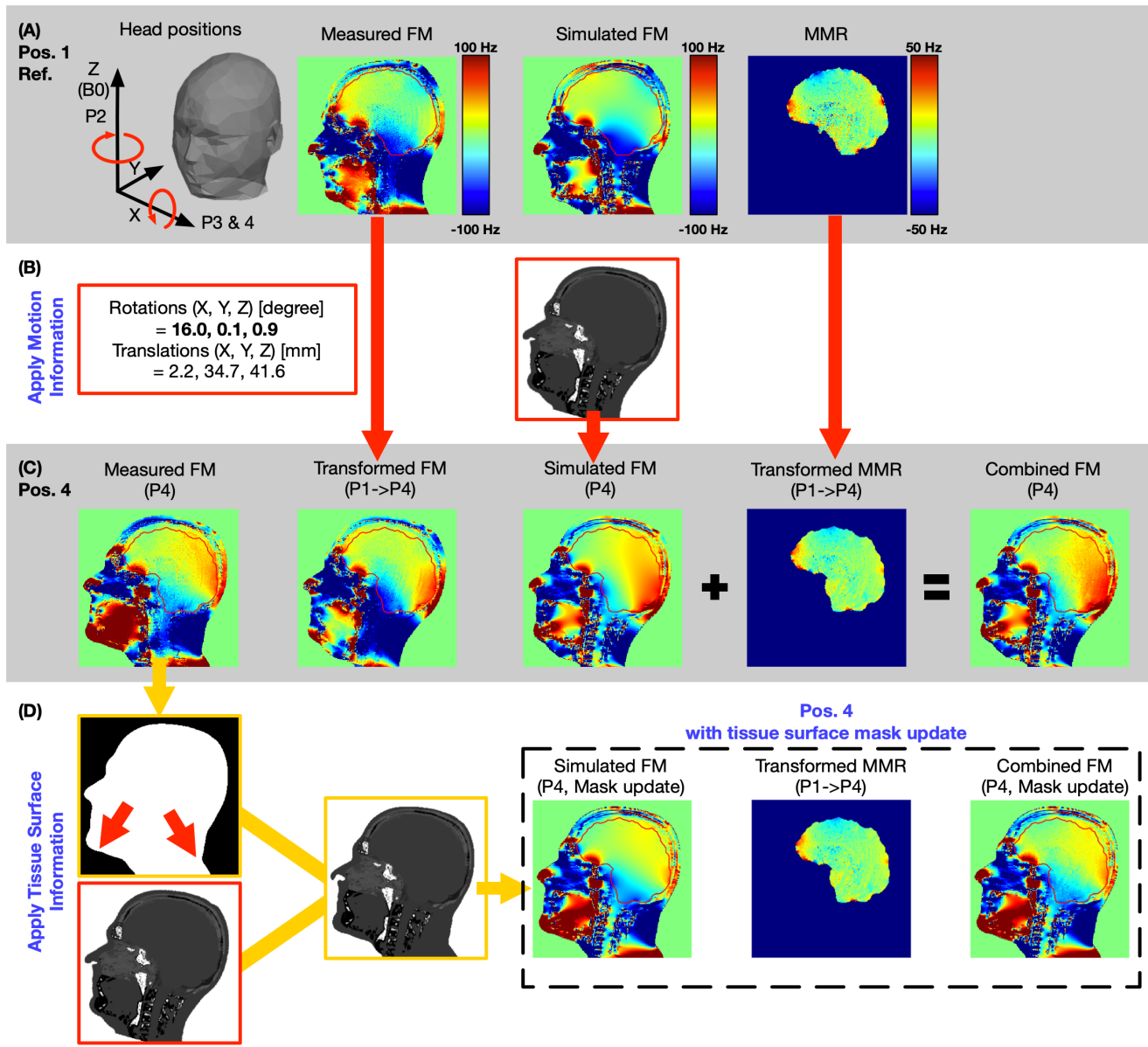

**Fig. 4** *Figure 4A-C illustrates the calculation of transformed FM, simulated FM, and combined FM from a reference position (Position 1) to a new head position (Position 4). The transform FM utilizes the measured FM in reference position with a simple rigid-body transform operation. The absolute rotation and translation values have been marked in red squares. The simulated FM is calculated from the forward B0 approximation, with a rigid-body transformed susceptibility map. The combined FM adds a rigid-body transform MMR field map from Position 1 into the simulated FM at a new head position. Figure 4C and 4D, show the process of susceptibility map mask adjustment when the large head rotation on X-axis causes subject tissue boundary changes. The updated head-mask is calculated from the measured field map at Position 4 and applied to the rigid body transformed susceptibility map. For large head movement, the simulated FM and combined FM were recalculated using the head-mask updated susceptibility map (dash square). The red arrow indicates the mask displacement.*

2.4 Head motions

$B_0$ was predicted for different head positions and compared with measurements in the new position. A a $\Delta B_0$ field simulation strategy (Simulated FM), a rigid-body transform strategy (transformed FM), and a combined strategy (Combined FM) were used to estimate motion-induced $B_0$ field variations (Figure 4A-C). Both 6-degrees of freedom (DOF) motion parameters and the rigid-body transform were obtained from an image-based retrospective correction method, FLIRT[28], in the FMRIB Software Library (FSL)[29] package.

For the rigid-body field-map transform strategy (Figure 4A-C), the calculated field map at the new head position was structured from the static SH shim field map and a rigid-body transformed measured field map was acquired at the reference position (Position 1). Note that the static SH shim field ($B_{SH}$) was first subtracted from the measured field map at the reference position in order to approximate the subject-induced $\Delta B_0$ variations. The rigid-body transform strategy is similar to a previously proposed template-based prediction method[30].

For the simulation strategy (Figure 4A-C), the field map at each new head position was approximated using the proposed simulation method as described in Method 2.2 (Equation 1). At each new head position, we assumed that the SH shim field ($B_{SH}$), phase error ($B_k$), and Chi field from the lower body ($B_{\chi'}$) remained unchanged since the head movement won't affect these factors. Accordingly, we need to calculate the Chi ($B_{\chi\_m}$) field from a position-dependent susceptibility model. The position-dependent susceptibility model was obtained by rigid-body transformation of the fitted susceptibility model at the reference position.

For the combined strategy (Figure 4A-C), a multiple model residual field map, MMR, which was the discrepancy between the measured and simulated field map at the reference head position (Position 1), was first rigid-body transformed and then added to the simulated field map at the new head position. The combined strategy is a simple derivative step of the simulation strategy. We included the transformed MMR into the simulation field map in order to mitigate the difference between the simulated field map and the measured field map at each new head position.

Large head motion around the X axis introduced a bulk $B_0$ off-resonance difference at the back of the head and neck (Figure 4C and D). This may have arisen due to the deformation of the subject's soft tissue geometry after a large head movement, which significantly changes the shape of the body boundaries (skin folds and stretches) in the neck and jaw region. To improve the prediction precision for large motion around X axis, a subject-specific geometry mask depicting the anatomy in this new position was calculated at Position 4 from the measured field map. Whereas the rest of the susceptibility model remains the same as calculated in the simulation strategy. Owing to an adjustment of the subject's geometry mask shape, both simulation and combined strategies were updated for large head.

2.5 Multi-coil array $B_0$ shimming simulation

To demonstrate the effectiveness of the proposed field map calculation methods, we compared the global shimming performance of measured field maps and the proposed calculated field maps.

We used a 16-channel multi-coil shimming array[10], the magnetic field from each shim coil was previously measured using a dual-echo GRE sequence (TE$_{1/2}$ = 2.8/7.8 ms, TR = 15 ms, FOV = 208*208*160 mm$^3$, FA = 12°) with 2mm isotropic resolution. Shim currents were estimated by the MATLAB function 'quadprog' to minimize a cost function[31], which was defined as the sum of the residual magnetic field:

$$cost\_s = \sum_{voxel} \left| B_0(\boldsymbol{r}) - B_0^{shim}(\boldsymbol{r}) \right|^2 \qquad (7)$$

where $B_0^{shim}(\boldsymbol{r})$ denotes the magnetic field created by shim coils.

In our global shimming simulation, the cost function was evaluated over all voxels in the brain after down sampling their size from 1mm to 2mm isotropic, the maximal shim current on each coil was limited to 4A, and the maximal total shim current across all shim coils was limited to 50A. The calculated shim current for each shim coil was then applied to the measured shim coil basis map to generate a counteracting magnetic field that counteracted brain-induced B$_0$ inhomogeneities.

During the simulation, we applied the calculated field maps to the cost function (Equation 6), to replicate the situation where a measured field map is not available at a new head position. The generated counter-acted magnetic field, estimated from the calculated field maps, was then used to shim the measured field map. For comparison, the measured field map was also shimmed. The shimming performance was evaluated by calculating the standard deviation of the measured field map before and after shimming (Vol. σB$_0$).

## 3. Results:

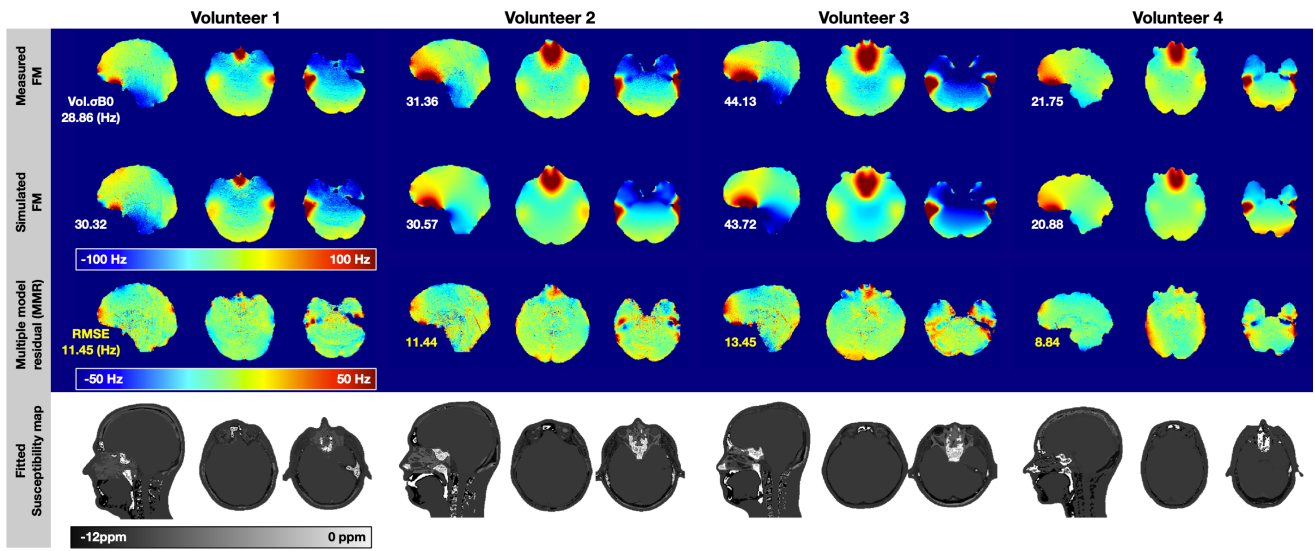

**Fig. 5** *Four volunteers simulation results at the reference position. First row: Measured field maps, color range [-100Hz, 100Hz]; Second row: Simulated field maps, color range [-100Hz, 100Hz]; Third row: Residual field maps (Measured - Simulated), color range [-50Hz, 50Hz]; Bottom row: Fitted susceptibility maps, color range [-12ppm, 0ppm].*

Figure 5 compares simulation performance across four subjects with the ground truth, measured field map. The measured field maps for three representative slices from four volunteers are shown in the first row. The simulated field maps are shown in the second row. The standard deviations (Vol.σ$B_0$) of the individual simulated field map and the measured field map have been listed on the left. The Vol.σ$B_0$ value for the simulated field map (ranging between 20.88Hz to 43.72Hz) is smaller to the measured field map (ranging between 21.75Hz and 44.13Hz). The RMSE value for the multiple model residual (Measured - Simulated) field map, MMR, was also listed on the third row. The residual field map has a different colormap limit (±50Hz V.S. ±100Hz), compared with measured and simulated field maps. As shown in the residual field map, the biggest discrepancies between the measured and simulated field maps were found around 1) prefrontal cortex region and temporal lobe region, which are close to the air-tissue interfaces, and 2) occipital lobe region, which is close to the superior sagittal sinus and transverse sinuses. The supplementary figure S3 summarized the residual field map improvement from using the 3-class model susceptibility map to the proposed fitted susceptibility map, in short, a 30% average improvement was found (from 16.06Hz to 11.29Hz). The bottom row shows the fitted susceptibility map. The sinuses and bony structure location are in good agreement with the anatomical UTE image.

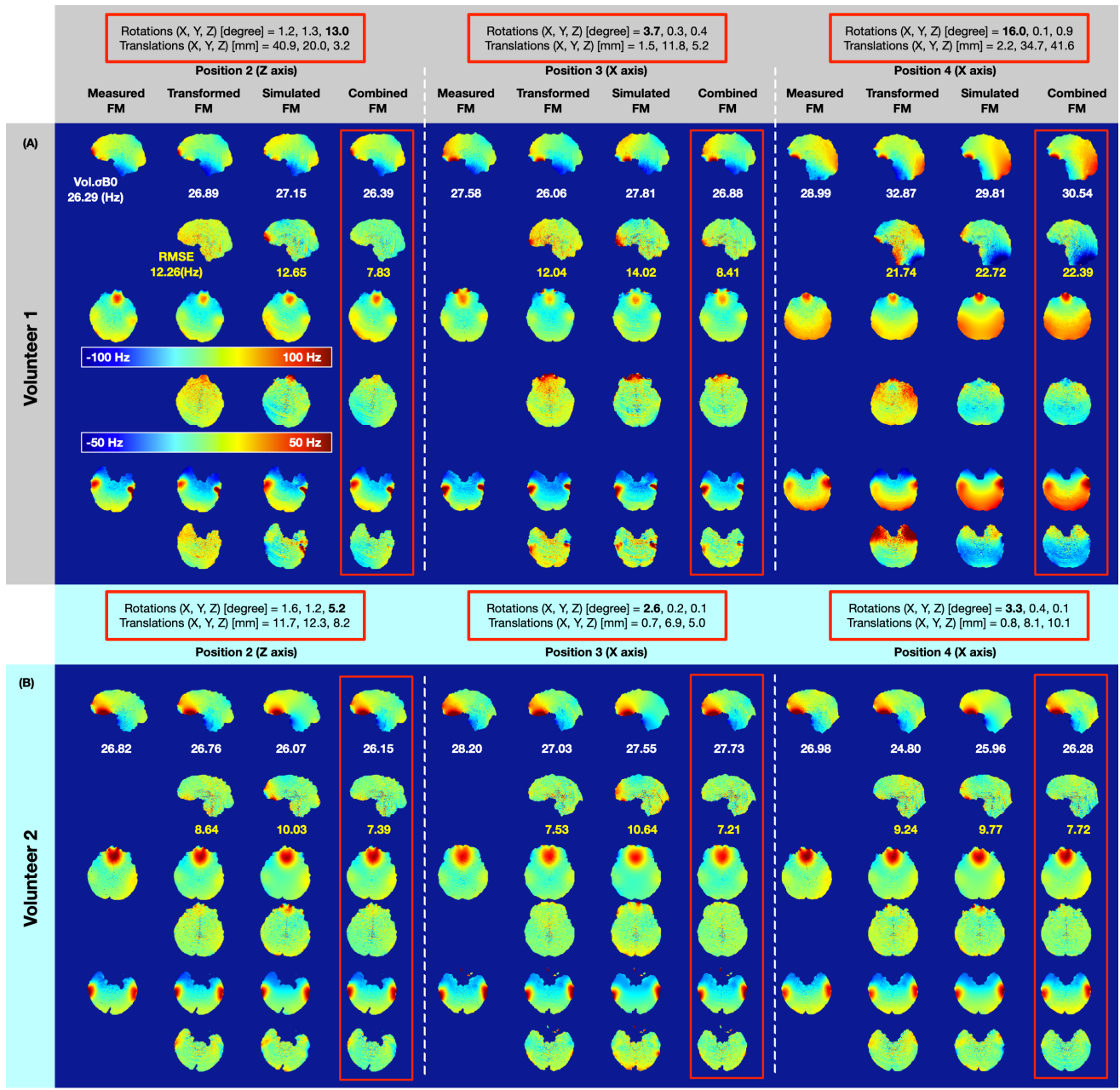

**Fig. 6** *Motion-induced field map estimation comparison in three representative slices from two subjects. The Vol. $\sigma B_0$ of the field map and RMSE of the estimated strategies were reported. The absolute rotation and translation values have been marked in red squares.*

Figure 6 compares the transformed FM, simulated FM, and combined FM with the measured FM in three representative axial slices from two volunteers. In figure 6A, volunteer 1 had rotated the head about the Z axis by 13 degrees. The Vol.$\sigma B_0$ values for the proposed field map estimation strategies are 26.89Hz, 27.15 Hz, and 28.39Hz for transformed FM, simulated FM, and combined FM, respectively. In comparison, the Vol.$\sigma B_0$ for a measured FM at Position2 is 26.29Hz (Figure 6A first column). The residual map between the measured and proposed strategy is shown below the field maps. The RMSE is 12.25Hz, 12.65Hz, and 7.83Hz for transformed FM, simulated FM, and combined FM, respectively. Compared with the simulated FM, the combined FM reduces the RMSE by 40% (from 12.65Hz to 7.83Hz).

In Position 3, there is roughly a 4-degree head rotation around the X axis. The Vol.σ$B_0$ for measured FM, transformed FM, simulated FM, and combined FM are 27.58Hz, 26.06Hz, 27.81Hz, and 26.88Hz, respectively. The RMSE of the combined FM (8.41Hz) was reduced by 40% with respect to the simulated FM (14.02Hz). The best performance was observed for Position 2, while the other positions had remaining field deviations, in the prefrontal cortex for all proposed strategies. This is because head rotation around the X axis usually causes a realignment of the air-tissue interfaces with respect to $B_0$. The significant $B_0$ change happens in the frontal sinus, which is far away from the center of the head compared to the ethmoid and sphenoid sinus. With even larger head rotation around the X axis, as shown in Position 4, the RMSE for the transformed FM, simulated FM, and combined FM is 21.74Hz, 22.72Hz, and 22.39Hz, respectively. Both estimation strategies have larger Vol. σ$B_0$ values compared to Position 2 and Position 3, where a bulk $B_0$ off-resonance difference was found at the posterior side. As shown in figure 6B, volunteer 2 had smaller head rotations in comparison to volunteer 1. Hence, the bulk $B_0$ off-resonance difference in position 4 was not found. Comparison of measured FM, transformed FM, simulated FM, and combined FM for all four volunteers are shown in supplementary table ST1.

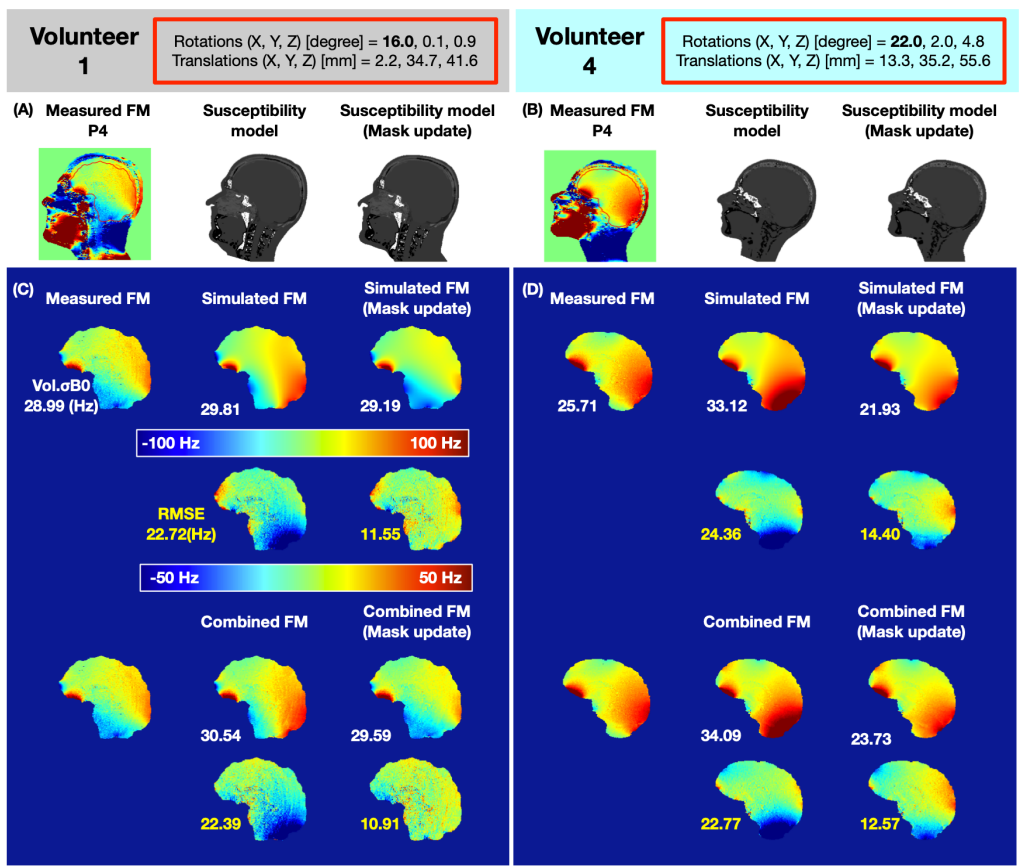

**Fig. 7** *Mask updated susceptibility models used for Position 4 in two volunteers. (A) Mask of volunteer 1 tissue boundary update process. (B) Mask of volunteer 2 tissue boundary update process. (C) Compare the simulation improvement (RMSE) by introducing the new subject tissue boundary for volunteer 1 (the second row vs. the third row). (D) Compare the simulation improvement (RMSE) by introducing the new subject tissue boundary for volunteer 2 (the*

*second row vs. the third row). The absolute rotation and translation values have been marked in red squares.*

In figure 6A, large head motion around the X axis introduced a bulk $B_0$ off-resonance difference at the back of the head and neck. This may have arisen due to the deformation of the subject's soft tissue geometry after a large head movement, which significantly changes the shape of the body boundaries (skin folds and stretches) in the neck and jaw region. Figure 7 shows improvement through simulated FM and combined FM after a large head movement owing to an adjustment of the subject's geometry mask shape. In figure 7A and B, a subject-specific geometry mask depicting the anatomy in this new position was calculated at Position 4 from the measured field map for volunteers 1 and 4, respectively. For volunteer 1, compared to the initial simulated FM without mask update in Figure 7C (in the middle column), the simulated FM with mask update has reduced the RMSE by 50% (from 22.72Hz to 11.55Hz). The combined FM with mask update has further reduced the RMSE by 52% (from 22.91Hz to 10.91Hz) in comparison to the initial combined FM. For volunteer 2, compared to the initial simulated FM without mask update in Figure 7D (in the middle column), the simulated FM with mask update has reduced the RMSE by 40% (from 24.36Hz to 14.40Hz). The combined FM with mask update has reduced the RMSE by 45% (from 22.77Hz to 12.57Hz) in comparison to the initial combined FM.

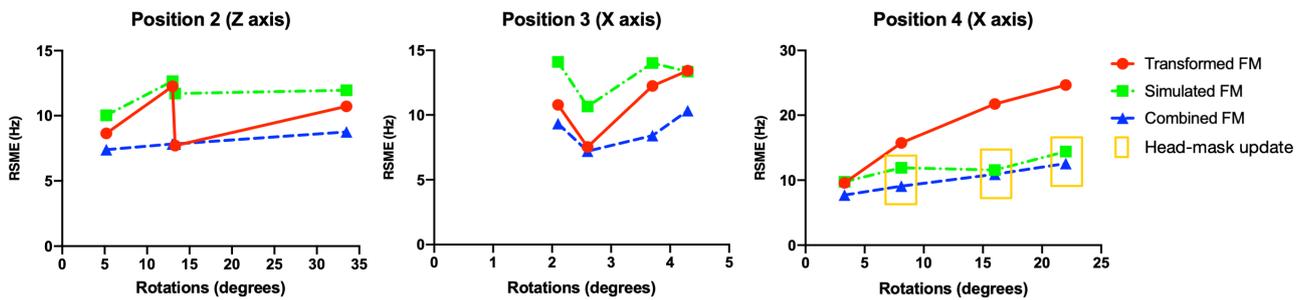

**Fig. 8** *Comparison of prediction performance between the rigid-body transform, the simulation, and the combined strategies at multiple head angles. The yellow window indicates the simulation and combined strategies with subject head-mask update.*

Figure 8 summarizes the rotation degrees in relation to the prediction performance for the rigid-body transform strategy, simulation strategy, and combined strategy at three head positions, for rotations around the z-axis, and for small and large rotations around the x-axis. For position 2, where rotations around the Z axis were recorded up to 33 degrees, both the rigid-body transform strategy and the combined strategy have similar prediction performance, with an average of 10Hz and 8Hz difference from the measured field map, respectively. For position 3, averaging about 10Hz and 8Hz differences from the measured field map were found for transform and combined strategies, respectively. However, the rotations around the X axis (Position 3) are usually less than 5 degrees. In comparison to transform FM, the combined FM reduced the RMSE to ground-truth measured field map by 18.3%. For position 4, with opposite rotations around the X axis, similar performance (about 10Hz difference from the measured field map) was for transform and combined strategies, when the rotation angle is below 5 degrees. Head motion with rotation above 5 degrees will normally change the shape of the anatomy shape substantially, therefore, the prediction performance for the rigid-body transform strategy degrades. Nevertheless, after updating the subject head-mask, the simulation and combined strategies retain a similar prediction performance as at small rotations (less than 5 degrees). Augmented with subject head-mask the combined FM reduced the RMSE to ground-truth measured field map by 25.8% (Transform FM V.S. Combined FM).

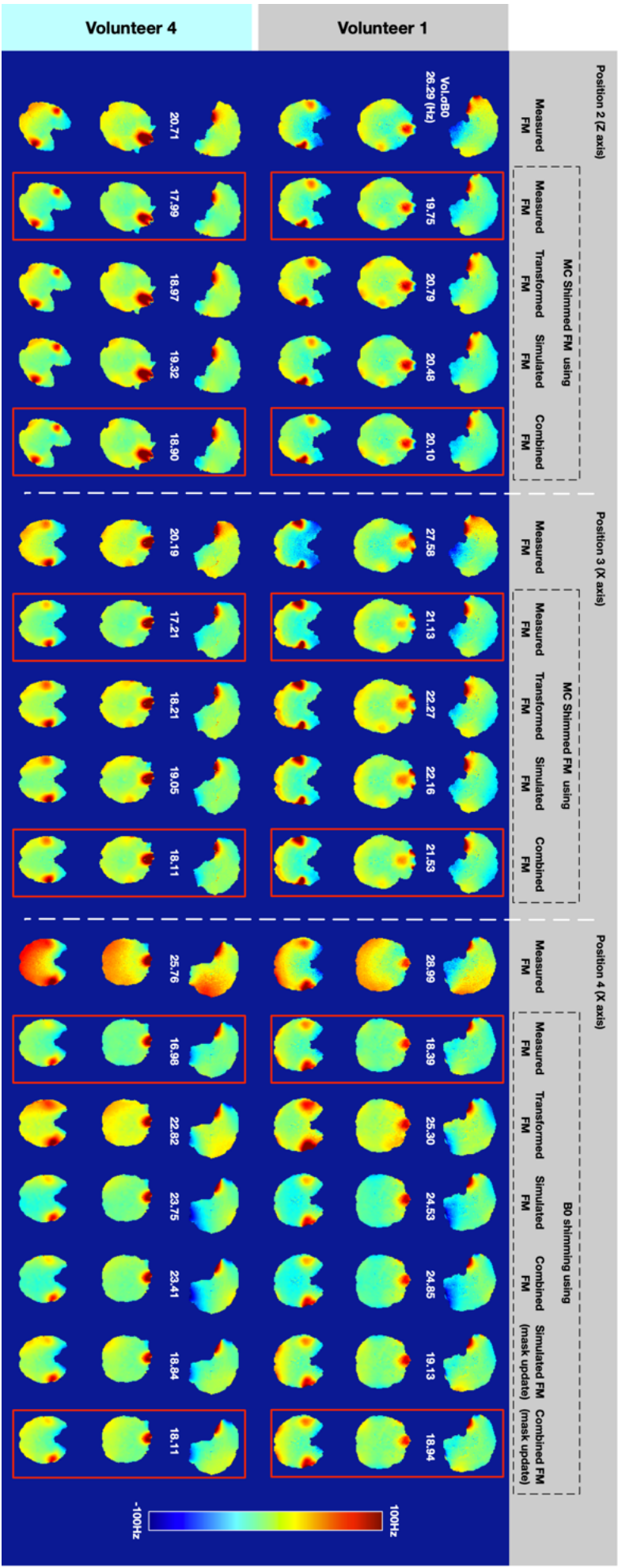

**Fig. 9** *Global shimming simulation off-resonance magnetic field maps at three representative slices from two volunteers. For positions 2 and 3, the first column is the target measured FM for B0 shimming. The second column to the fifth column compared the residual field map after B0 shimming with proposed strategies, where the second column used the measured FM itself as the baseline. For position 4, additional B0 shimming results using the mask update susceptibility model is compared with the B0 shimming result using measured field FM.*

Figure 9 shows the global shimming performance of the measured field map using proposed field map calculation strategies as a substitute. In each head position, the first column shows the measured field map as the $B_0$ shimming target. The shim coil setup, as well as simulation parameters, are shown in supplementary figure S4A. For the $B_0$ shimming experiment, the shim performance results are shown in the second to fifth columns. We used the shim result by input measured field map as the baseline. For position 2 of volunteer 1, global $B_0$ shimming with a measured field map could reduce Vol.$\sigma B_0$ from 26.29Hz to 19.75Hz. The shim performance using the rigid-body transform FM, simulation FM, and combined FM as a $B_0$ shimming target reduce the Vol.$\sigma B_0$ of the measured field map to 20.79Hz, 20.48Hz, and 20.10Hz, respectively. The Vol.$\sigma B_0$ of residual field map using the proposed field map calculation strategies are similar, with the combined strategy slightly close to the baseline. The spatial distributions of the residual field map after global shimming using combined FM are comparable to the one shimmed with the measured FM. In position 4 of both volunteers 1 and 4, the head rotation is relatively large, and the Vol.$\sigma B_0$ of residual field map using proposed rigid-body transform FM, simulation FM, and combined FM is 25.30Hz, 24.53Hz, and 24.85Hz respectively, whereas the baseline is 18.39Hz. The bulk $B_0$ off-resonance difference at the back of the head and neck (as shown in figure 6 and figure 7) leads to an incorrect $B_0$ shimming calculation. With the adjustment of the subject's geometry mask shape, in the case of volunteer 1, the Vol.$\sigma B_0$ of residual field map using simulated FM and combined FM with mask update has reduced by 22% (from 24.53Hz to 19.13Hz) and 23% (24.85Hz to 18.94Hz), receptively. For volunteer2, the Vol.$\sigma B_0$ of residual field map using simulated FM and combined FM with mask update has reduced by 2% (from 23.75Hz to 18.84Hz) and 23% (23.14Hz to 18.11Hz), receptively. Supplementary figure S4B provides simulation results for all four subjects.

## 4. Discussion:
In this work, we proposed a $\Delta B_0$ field simulation strategy using a fitted subject-specific susceptibility model constituted from a UTE image and a reference field map. We verified the prediction error of our approach to validate it for the case that an experimentally measured field map is not available. Compared with the rigid-body transformed field map, the proposed simulation strategy based on local susceptibility effects and position-specific masking achieved superior prediction performance at multiple head positions.

Substantial discrepancies in these methods were found when predicting motion-induced $B_0$ changes with a large head motion around the X axis. However, with additional information from subject anatomy in the new head position, the simulation strategy and combined strategy were able to mitigate this discrepancy.

For the proposed simulation strategy, there are a few considerations that determine the accuracy of the simulated field map:

4.1 The Fourier approximation method

We adopted the Fourier based dipole-approximations method[17, 18], which enables rapid computation of the magnetic field, using a high sampling resolution and spatial padding factor. The spatial padding extends the computational volume outside the imaged FOV, to make sure that the perturbing induction fields have diminished at the computational volume boundaries, avoiding artifacts like folding over of the field into the opposite computational volume side. This is primarily due to the constraints on the k-space scaling factor in Equation. 2 which is singular at $k$ = 0. along with the periodic nature of the Discrete Fourier Transformation (DFT). It is also important to stress that low-resolution sampling limits computational accuracy through misrepresentations of the compartment borders as well as the k-space scaling coefficient (i.e., insufficient sampling of the dipole response in k-space due to low grid resolution) in Equation. 2[19, 32].

Based on previous research[21, 32], the pitfalls of the Fourier dipole approximation method could be carefully mitigated through high spatial sampling (with voxel size being less than 1mm isotropic) and the spatial padding factor (greater than 3).

4.2 The measured field map

We used the measured field maps as a ground truth $B_0$ map for the proposed simulation. However, it is difficult to acquire an unbiased $B_0$ map in the brain, particularly in voxels near the air-tissue interface where signal loss from intra-voxel dephasing and extreme phase wrapping behavior make accurate $B_0$ measurements very challenging. In addition to that, the phase offset caused by eddy currents is another source of disagreement between the measured and simulated field maps. Compared to vendor-provided dual-echo 2D GRE, we used a dual-echo 3D GRE sequence for field mapping. We tried to minimize the readout and slice-selective gradient eddy currents by using monopolar readout and non-selective excitation. Unfortunately, several factors can still influence the phase errors in the field mapping sequence, particularly in the phase encoding direction, inducing 1) eddy currents in the phase-encoding direction, 2) a disagreement between simulated and actual SH shim field, 3) gradient-offsets due to the ensuing FOV shifts from the iso-center.

In the present study, we observed a linear field map contribution (from anterior to posterior, the phase-encoding direction) within the measured field map, that was present after demodulating the simulated SH shim field and sample-induced $B_\chi$ field (Figure 3). It causes a shift of the fitted dipole source towards the posterior direction which would then fall outside the body region - a physiologically not meaningful result. To address this issue, we used k-space shifts modeling to estimate the residual field to its first order, an idea previously explored by Diefenbach et al[33]. The introduction of a phase gradient error, approximated by a linear field, diminishes the discrepancies between the measured and simulated field map.

### 4.3 Tissue susceptibility models

The accuracy of the subject-specific susceptibility model for discriminating air-bone boundaries plays a critical role for the quality of our field map simulations. The temporal and the nasal region are composed of complex structures comprising soft tissue, mucus, air, and cartilage, which usually result in strong partial-volume effects. It is challenging to fully resolve the different tissue types with the spatial resolution used in this work. To circumvent tissue misclassification, we propose an iterative, nonlinear optimization algorithm to determine susceptibility values for each segment of the tissue map obtained from UTE. The optimization algorithm used the residual error within the brain mask, obtained from the first three-class tissue field map. The proposed iterative nonlinear optimization algorithm is similar to a previously published phase replacement method[23], which is used to determine the susceptibility values of air, bone, and teeth. Both algorithms require a segmented air and bone map and phase (frequency) information around the structure of interest. Compared with the phase replacement method, where the segmented air and bone masks were obtained from a short echo GRE image, our multiple segment model is based on the inverse log-scaled UTE image, which facilitates identification of air-tissue and air-bone boundaries.

In contrast to a previously published method[21] which requires anatomical data with CT and MRI, the use of a UTE sequence provides adequate anatomical information for segmentation of air/bone boundaries in the human head. As a result, the need for CT images to detect bone boundaries can be addressed with a MR scanner. The UTE sequence gives an analogous histogram distribution to the inverse log-scaled ZTE images, introduced by Wiesinger et. al[25]. We use a flip-angle of 1 degree to suppress signals from both water and fat to provide a uniform soft-tissue contrast (PD-like) and therefore detection of the bone. The low-flip angle (smaller than the Ernst angle) also avoids T1 saturation of long T1 components (e.g., CSF and eyes) that can lead to misclassification as bony structures. Compared with previously proposed UTE methods[34, 35] using T2 relaxation differences, the short TE and low flip-angle UTE take advantage of PD contrast and do not require long T2 or T1 suppression, like echo subtraction or inversion pulses.

The UTE intensity taken on its to obtain a multi segment tissue model is not sufficient to obtain a high quality field map estimation. This is shown in supplementary figure S5B, where results for fitting the susceptibility using 34 or 330 UTE-based tissue classes are shown. This happens because the actual susceptibility can differ between head regions, although the tissues still have similar UTE intensities. For instance, the mastoid cavity was reported to have lower susceptibilities than the rest of the sinuses[23]. In principle, the susceptibility value could be determined for each individual voxel within the multiple segment model. However, estimating the susceptibility value voxel-by-voxel is time consuming and may be impractical. Therefore, we subdivided the head into several VOIs, and obtained VOI-specific estimates of the susceptibility for each tissue class.

Without spatially defined VOIs, the field map error was only reduced by 3% (from 15.91Hz to 15.56Hz), whereas a 34 tissue class model with 4 spatially defined VOIs could reduce the residual error by 20% (from 15.91Hz to 12.60Hz).

The proposed iterative nonlinear optimization algorithm is able to approximate the susceptibilities for various tissue classes, however, the approximation accuracy at this stage is still rudimentary in comparison to the previously published phase replacement method[23]. Further investigations, including a comprehensive frequency map for the fitted susceptibility model (field map of the whole head, instead of the brain), are needed to further improve approximation accuracy for future applications.

4.4 Motion induced field map variations

As shown in Figure 6 and Figure 7, estimating a field map at a new head position using the rigid-body transform strategy and corresponding motion parameters (6 DOFs) from image-based retrospective motion correction provides a simple and good approximation for small motion-induced $\Delta B_0$ variants (<5 degrees). Nonetheless, the rigid-body transform strategy is insufficient to predict the brain field map due to large motion. The field map error becomes particularly large in the jaw and neck regions. Our extended model showed that $\Delta B_0$ shifts arise from the displacement of the jaw and neck, leading to alterations of the $\Delta B_0$ field in the brain, especially in its inferior regions. Compared to the rigid-body transform strategy, the combined field map strategy has a similar performance in predicting the field map for small movements. Furthermore, the image-based retrospective motion correction method requires a series of volumes and is thereby limited in terms of temporal resolution. In future work, we plan to combine the simulation strategy with a motion capturing camera system[36], in order to be able to dynamically predict the motion-induced $\Delta B_0$ variants without the need for further MR measurements. This would also allow for applications like dynamic $B_0$-shimming and correction during EPI time courses or high resolution $T_2^*$-mapping independent of the temporal resolution of the imaging sequence.

In conclusion, we found that a rigid-body transformation of a measured field map provides a feasible approximation for small motion-induced $\Delta B_0$ variants. The proposed field map simulation strategy can estimate the motion-induced $\Delta B_0$ changes without MR measurement. A further improvement was observed by taking the actual position and subject geometry into account. This modeling strategy may be used for improved image reconstruction, better (dynamic) $B_0$ shimming and $B_0$ related artifacts correction in cases where no measured field map is available for the different head positions.

**Acknowledgments**
This work was funded in part by DFG, a Reinhard Koselleck Project, DFG SCHE 658/12, DFG SCHE 658/13, ERC advanced grant No 834940, and by the Max Planck Society.

**Author contributions**
JZ study conception and design, acquisition of data, analysis and interpretation of data, drafting of manuscript. GH study conception and design, acquisition of data, analysis and interpretation of data, critical revision. AA study conception and design, critical revision. JB study conception and design, critical revision. MZ study conception and design, critical revision. KS study conception and design, critical revision.